\documentclass[10pt,twocolumn,superscriptaddress,sort&compress,floatfix]{article}

\usepackage{geometry}
\usepackage[numbers,sort&compress]{natbib}  
\usepackage{balance}
\usepackage{graphicx}
\usepackage{amsmath}

\newcommand{\anappendix}[1]{Appendix~#1}

\begin{document}

\twocolumn[{

\begin{raggedright}
\Large \textbf{Polymer compaction and bridging-induced clustering of protein-inspired patchy particles}
\par
\end{raggedright}

\vspace{0.4cm}

{\large C.~A. Brackley}

{\scriptsize SUPA, School of Physics \& Astronomy, University of Edinburgh, Peter Guthrie Tait Road, Edinburgh, EH9 3FD, UK}

\vspace{0.4cm}

\textbf{Abstract}\vspace{0.1cm}

There are many proteins or protein complexes which have multiple DNA binding domains. This allows them to bind to multiple points on a DNA molecule (or chromatin fibre) at the same time. There are also many proteins which have been found to be able to compact DNA \textit{in vitro}, and many others have been observed in foci or puncta when fluorescently labelled and imaged \textit{in vivo}. In this work we study, using coarse-grained Langevin dynamics simulations, the compaction of polymers by simple model proteins and a phenomenon known as the ``bridging-induced attraction''. The latter is a mechanism observed in previous simulations [Brackley \textit{et al.}, \textit{Proc. Natl. Acad. Sci. USA} \textbf{110} (2013)], where proteins modelled as spheres form clusters via their multivalent interactions with a polymer, even in the absence of any explicit protein-protein attractive interactions. Here we extend this concept to consider more detailed model proteins, represented as simple ``patchy particles'' interacting with a semi-flexible bead-and-spring polymer. We find that both the compacting ability and the effect of the bridging-induced attraction depend on the valence of the model proteins. These effects also depend on the shape of the protein, which determines its ability to form bridges.

\vspace{0.3cm}


\vspace{0.6cm}

}]

\section*{Introduction}

\vspace{-5mm}

Within both bacteria and eukaryotes, there are many DNA binding proteins which interact with the genome to perform different functions. Many of these proteins form complexes which have more than one DNA binding domain. Such a complex can then bind the chromosome at multiple places at the same time, forming molecular bridges. This is thought to be at the basis of several mechanisms for gene regulation. For example, in higher eukaryotes gene expression from a promoter is often activated when proteins (i.e., transcription factors, TFs) bind at a site on the chromosome called an enhancer; promoters and enhancers can be separated by millions of base-pairs of DNA~\cite{Albertsbook}. The current model for how enhancers operate is that they must come into close physical contact with the promoter, forming a chromosome loop, and that this interaction is mediated by a complex of DNA binding proteins (e.g., TFs and RNA Polymerase II). Another regulatory mechanism is that chromosome regions which need to be inactivated are often compacted---again bridge-forming DNA binding complexes have been implicated (e.g., the heterochromatin protein HP1~\cite{Verschure2005}, or the polycomb repressive complexes, PRC1 and PRC2~\cite{Eskeland2010}).

Previously~\cite{Brackley2013,Johnson2015}, using simple polymer simulations, we uncovered a remarkable property of bridge forming proteins interacting with DNA or chromatin (the fibrous material formed from DNA and proteins which makes up eukaryotic chromosomes~\cite{Parmar2019}). When protein complexes are represented in a simulation as simple spheres with an isotropic attractive interaction with the polymer, they will spontaneously form clusters on that polymer, even in the absence of any explicit protein-protein attraction [Fig.~1(a)]. This occurs because once a protein binds to two polymer regions to form a bridge, there is a local increase in polymer density---a second protein is more likely to bind the polymer in the same region. This sets up a positive feedback (bridging increases local polymer density which increases bridging). We termed this effect the \textit{bridging-induced attraction} (BIA). An alternative way of looking at the effect is that forming a polymer-protein-polymer bridge leads to an enthalpy gain, but an entropic loss (a bridge forms a loop, which reduces the entropy of the polymer). Placing a second bridge in the same region gives a second enthalpy gain without incurring a further entropic penalty. 

\begin{figure}[ht!]
\centering
\includegraphics[width=0.48\textwidth]{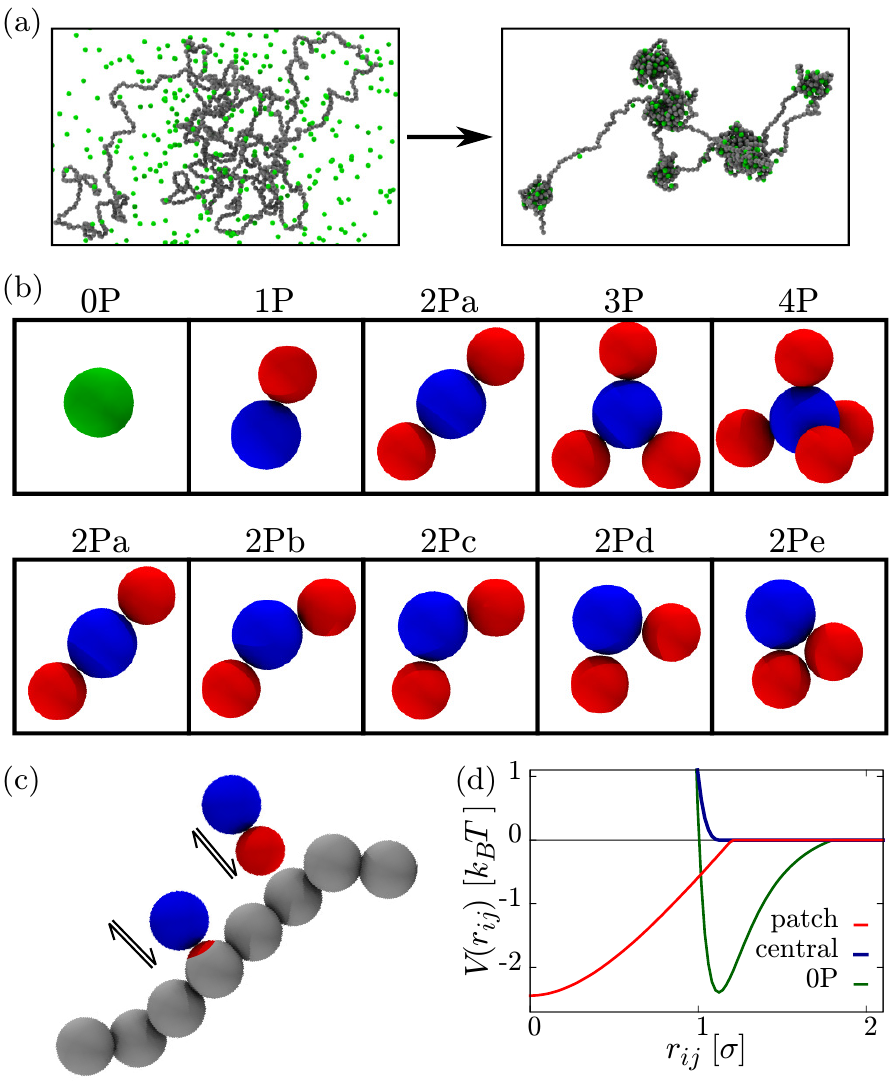}
\caption{The bridging-induced attraction (BIA). (a) Snapshots of a simulation of isotropic spheres interacting uniformly with a polymer. The initial condition is shown on the left; the BIA leads to clustering (right; snapshot after $10^5\tau$---at longer times spheres merge into a single cluster). For clarity, in the right hand image only spheres bound to the polymer are shown (a sphere is said to be bound to a polymer bead if their centre-to-centre distance is less than $1.7~\sigma$). (b) Schematic showing the patchy model proteins. Top row: increasing number of patches (valence) which are spaced equidistantly. Bottom row: two patches are positioned at different angles. (c) Schematic showing how patches (red) interact with polymer beads via a purely attractive interaction, while the central bead (blue) interacts with purely repulsive WCA. For the former, the energy minimum is such that the patch sits inside the polymer bead. (d) Plot showing the functional form of the protein-polymer interaction potential for isotropic spheres (0P model; green), the central bead of patchy proteins (blue), and the patches (red).  }
\end{figure}

More specifically, the BIA only operates when the proteins can form bridges (i.e., they can bind at least two polymer regions at a time), and the interaction energy must be strong enough such that bridges form for a sufficiently long time for cluster growth to commence. We previously~\cite{Brackley2013} found that the mechanism is robust to changes in protein size and polymer stiffness (though this can alter the shape of the emerging clusters). We also previously found that the mechanism still works for the minimal case of proteins which can only bind two polymer regions at the same time, though the effect is weaker. Here, we investigate the effect in more detail with regard to the valency and shape of the model proteins. We also examine the ability of these model proteins to compact the polymer, which was also previously studied for the case of isotropic spheres~\cite{Nicodemi2009,Barbieri16173}

In other previous work, we found that by ``patterning'' the polymer with specific (strong attractive interaction) and non-specific (weak attractive interaction) protein binding sites, the action of the BIA drives the polymer into specific structures~\cite{Brackley2013,BrackleyNAR}. By introducing multiple species of spherical proteins with different binding sites, simulations can reproduce the structure observed in mammalian genomes~\cite{BrackleyNAR}. This includes the ``compartmentalisation'', or spatial segregation of active and inactive regions of the genome, and the formation of domains of enriched self-interactions (often called ``topologically associated domains'', or TADs, in the chromatin biology literature)~\cite{Bouwman2015}. For example, with just two species of protein, and with two sets of binding sites inferred from experimental data, simulations can correctly predict over 80\% of the positions of TAD boundaries in human chromosomes (as revealed in chromosome-conformation-capture experiments~\cite{Schmitt2016}). We also showed that for non-specific binding of isotropic sphere model proteins with an excess of polymer, the clusters will grow, or coarsen, until there is a single large cluster containing protein and polymer beads, surrounded by a region with unbound polymer and low protein density~\cite{BrackleyEphemeral}. In other words the proteins proceed to phase separate in a way reminiscent of a binary fluid~\cite{Thomson1987,cates_tjhung_2018}. The coarsening can be arrested either by introducing specific strong binding sites, or by non-equilibrium processes~\cite{BrackleyEphemeral}.

Here we study the BIA in the context of more detailed model proteins. Specifically, we perform Langevin dynamics simulations of simple ``patchy'' model proteins with a finite number of polymer binding sites. In the next section we give details of the model and simulations. We then present results for model proteins with valence (number of ``patches'') 2, 3 or 4, both for the case where there is an abundance of polymer and low protein concentration (studying cluster formation via the BIA), and where there is high protein concentration such that the ``binding sites'' on the polymer become saturated (and we examine the resulting polymer compaction). Finally we present a similar analysis for the case of valence-2 proteins with patches at different angles---i.e., we examine the effect of protein ``shape''. Our main results are that the BIA operates more strongly for higher valence proteins, and that the main contribution of protein shape is that it affects the ability of the proteins to form bridges. This will be of interest in understanding the mechanisms through which DNA-binding proteins form foci in cells, and how they impact on genome organisation and regulation.

\vspace{-3mm}

\section*{Model}

\vspace{-2mm}

To study the BIA in more detail, here we use Langevin dynamics simulations of a simple polymer model, with proteins represented by ``patchy particles'' [Fig.~\ref1(b)]. The latter are rigid bodies made up from spheres (a ``central'' bead surrounded by ``patch'' beads), similar to the patchy particles models common in the soft matter physics literature~\cite{Zhang2004,Bianchi2008,Fusco2013,Teixeira2017}. We use a common bead-and-spring polymer model~\cite{Kremer1990}, which could represent DNA or chromatin---we use a persistence length close to that of the latter. $L$ beads of diameter $\sigma$ are connected in a chain via FENE bonds, described by the potential
\[
V_{\rm FENE}(r_{i,i+1}) = - \frac{K_{\rm FENE} R_0^2}{2} \log \left[ 1 - \left(\frac{r_{i,i+1}}{R_0}\right)^2 \right],
\]
where $r_{i,i+1}=|\mathbf{r}_{i+1}-\mathbf{r}_i|$ is the separation between two beads in the chain, $k_BT$ is the Boltzmann constant multiplied by the temperature, and $K_{\rm FENE}=30~k_BT$ and $R_0=1.6~\sigma$ are an energy scale and the maximum bond length respectively. Polymer bending rigidity is provided by a Kratky-Porod potential given by
\begin{equation}\label{eq:Vbend}
V_{\rm BEND}(\theta_i) = K_{\rm BEND} \left[ 1+\cos(\theta_i)\right], 
\end{equation}
where $\theta_i$ is the angle formed between three consecutive beads along the chain, defined by $\cos(\theta_i)=\mathbf{\hat{t}}_i\cdot\mathbf{\hat{t}}_{i+1}$, where
\[
\mathbf{\hat{t}}_i = \frac{ \mathbf{r}_{i+1}-\mathbf{r}_i  }{|\mathbf{r}_{i+1}-\mathbf{r}_i| }.
\]
$K_{\rm BEND}$ is the bending rigidity, related to the intrinsic polymer persistence length $l_p= K_{\rm BEND}\sigma/(k_BT)$; we set $l_p=3~\sigma$. Finally, steric interactions between polymer beads is provided by a Weeks-Chandler-Anderson (WCA) potential given by
\begin{equation}\label{eq:WCA}
V_{\rm WCA}(r_{ij}) = 1 + 4 \left[ \left( \frac{\sigma}{r_{ij}} \right)^{12} -  \left( \frac{\sigma}{r_{ij}} \right)^6  \right] ,
\end{equation}
where $r_{ij}=|\mathbf{r}_{i}-\mathbf{r}_j|$ is the separation between any two beads $i$ and $j$.

We consider nine different patchy protein models [Fig.~1(b)]. Previous simulations of patchy particles have often involved a complicated potential where interactions are determined by the angle between points on the surface of a spherical bead (e.g.,~\cite{Rovigatti2018}); here, we instead use a simple scheme where the central bead interacts sterically with polymer beads, while patch beads interact with the polymer via an attractive interaction but no steric interaction. The minimum of the potential is where the patch bead-polymer bead separation is zero, such that when ``binding'' a patch sits inside a polymer bead [Fig.~1(c)]. By using a short range interaction, in practice when it is bound  a patch either sits inside one polymer bead, or sits between two consecutive beads along the chain ($i$ and $i+1$). As detailed in \anappendix{A} we check that in each simulation no single patch interacts with more than one non-consecutive polymer bead (i.e., a single patch cannot form a bridge). 

More specifically, protein central beads have diameter $\sigma$ and interact with polymer beads and all other protein beads and patches via the WCA potential given in Eq.~(\ref{eq:WCA}). Patch beads (also diameter $\sigma$) interact with central and other patch beads via the WCA, but interact with polymer beads via a potential
\begin{align}
V_{\rm PATCH}(r_{ij}) =& \epsilon_0 [ (e^{-2ar_{ij}} - 2e^{-ar_{ij}}) \nonumber \\
&- (e^{-2ar_{\rm c}} - 2e^{-ar_{\rm c}}) ], \label{eq:Vpatch}
\end{align}
for $r_{ij}<r_{\rm c}$ and $V_{\rm PATCH}(r_{ij}) = 0$ otherwise. Here $r_{ij}$ is the separation between patch $i$ and polymer bead $j$, $a$ is a length set at $a=1.75~\sigma$, and the range was set at $r_{\rm c}=0.9~\sigma$. This has a similar functional form to the commonly used Morse potential. The energy $\epsilon_0$ determines the interaction strength; we note that the minimum energy of this potential (when $r_{ij}=0$) is not $\epsilon_0$ due to the terms in $r_c$. For clarity in plots we use the actual binding energy $\epsilon_P$; since it is possible for a protein patch to sit between two consecutive polymer beads, this energy can be a sum of contributions. As detailed in \anappendix{A}, to ensure that we plot the correct value of $\epsilon_P$, we measure it from the simulations for each $\epsilon_0$ value.
These potentials are plotted in Fig.~1(d). In summary, there are attractive protein-polymer interactions with strength $\epsilon_P$, but importantly there are no attractive protein-protein interactions.

We name the different models ``1P'', ``3P'', etc. for proteins which have 1 patch, 3 patches and so on. We also consider five varieties of protein with two patches named ``2Pa'' to ``2Pe'', which have patches separated by angles $\pi,0.825\pi,0.65\pi,0.475\pi$, and $0.3\pi$ respectively. As a comparison we also ran simulations with isotropically interacting spheres (e.g., as in Refs.~\cite{Brackley2013,BrackleyNAR}, here denoted the 0P model), where the protein-polymer interaction was determined by a shifted and truncated Lennard-Jones potential given by 
\begin{eqnarray}
V_{\rm LJ,cut}(r_{ij}) =& 4\epsilon \left[ \left( \frac{\sigma}{r_{ij}} \right)^{12} - \left( \frac{\sigma}{r_{ij}} \right)^{6} \right] \nonumber\\ 
&- 4\epsilon \left[ \left( \frac{\sigma}{r_{c}} \right)^{12} - \left( \frac{\sigma}{r_{c}} \right)^{6} \right] \nonumber
\end{eqnarray}
for $r_{ij}<r_{\rm c}$ and $V_{\rm LJ,cut}(r_{ij}) = 0$ otherwise; $r_{ij}$ is the separation between sphere protein $i$ and polymer bead $j$. We set $r_c=1.8~\sigma$. The isotropic spheres can can interact with as many polymer beads as allowed  sterically. Another difference between this and the patchy proteins case is that in the latter, on the whole, a polymer bead can interact with at most one protein patch at at time. This is because when bound, patch beads sit inside the polymer beads; since they interact with other patches through a WCA potential this will exclude another patch from interacting with the same polymer bead. In the isotropic 0P case a polymer bead can interact with multiple proteins.

A common approach in Langevin dynamics is to employ periodic boundary conditions; however, unless a sufficiently large enough system is used, this allows configurations where the polymer interacts with itself (through the bridging proteins) across the periodic boundaries. To avoid such configurations, and to avoid having to perform large simulations with a large number of proteins (to achieve concentrations sufficient for the BIA to act), here we instead confined all beads within a cubic region of width $l_x$. This is achieved through a ``wall'' potential $V_{\rm WALL}(r_{ij})$, where here $r_{ij}$ is the separation between bead $i$ and the closest point to it on wall $j$. Six such walls give the faces of the cube, and the potential takes the value 
\[
V_{\rm WALL}(r_{ij}) = -K_{\rm WALL} r_{ij}^2,
\]
if the bead is on the ``outside'' of the wall or 0 otherwise. We set $K_{\rm WALL}=10~k_BT$. As detailed for each set of simulations, typically we choose $l_x$ to be slightly larger than the mean radius of gyration for the polymer as predicted by the worm-like-chain (WLC) model~\cite{RubinsteinBook}. This means that the polymer is under (weak) confinement, and so has a lower entropy than it would in the absence of confinement; nevertheless, the polymer concentration is still much lower than one would typically find for chromatin or DNA \textit{in vivo}.

We performed Langevin dynamics simulations using the LAMMPS molecular dynamics software~\cite{Plimpton1995}. Briefly, the position of polymer bead $i$ evolves according to the equation
\[
m\frac{d^2\mathbf{r}_i}{dt^2} = -\boldsymbol{\nabla} V_i -\xi \frac{d\mathbf{r}_i}{dt} + \sqrt{2k_BT\xi} \boldsymbol{\eta_i}(t),
\]
where $m$ and $\xi$ are the mass and friction respectively, $V_i$ is the sum of interaction potentials for bead $i$, and $\boldsymbol{\eta}_i(t)$ is a noise term with components where  $\langle \eta_{i\alpha}(t) \rangle=0$ and $\langle \eta_{i\alpha}(t)\eta_{i\beta}(t') \rangle= \delta_{\alpha,\beta}\delta(t-t')$. The centre of mass of each rigid body patchy protein moves according to the same equation, with its orientation governed by a similar equation for rotation~\cite{Plimpton1995}. For simplicity we set the mass and friction the same for each polymer bead and each protein component bead.

The simulation length, mass, and energy units, $\sigma$, $m$, and $k_BT$ give rise to a time unit $\tau=\sqrt{m\sigma^2/k_BT}$. The friction coefficient is related to the diffusion constant via the Einstein relation $D=k_BT/\xi$, and we set $\xi=2$ which gives over-damped motion with $D=0.5~\sigma^2\tau^{-1}$ leading to a Brownian time of $\tau_B=\sqrt{\sigma^2/D}=2~\tau$. To map these units to a real physical system one could, e.g., assume that for chromatin $\sigma\approx30$~nm, $T$=300~K, and the nucleoplasm viscosity $\nu=10^{-2}$~Pa$\,$s (where from the Stokes equation for a sphere of diameter $d$, $\xi=6\pi\nu d$); this then gives $\tau=0.3$~ms.

In LAMMPS the Langevin equations are integrated using a velocity-Verlet algorithm; we set the time step dt$=0.01\tau$. To initialise simulations, proteins were positioned randomly, while polymer beads were placed along a random walk. Initialisation simulations were carried out using ``soft'' interaction potentials (to remove bead overlaps without numerical instabilities), before these were changed to the interaction potentials detailed above. Each simulation was then further equilibrated by running for $5\times10^4~\tau$ in the absence of attractive interactions. Simulations were then run \textit{with} protein-polymer attraction for at least $10^5~\tau$, by which time all measured quantities had stopped systematically changing. Values shown in figures are obtained from an average over configurations taken at regular intervals from a further simulation run of $10^5~\tau$; we also average over 5 independent simulations. In plots the error bars show the standard error in the mean.

\vspace{-3mm}

\section*{Results}

\vspace{-2mm}

\begin{figure*}[ht!]
\centering
\includegraphics{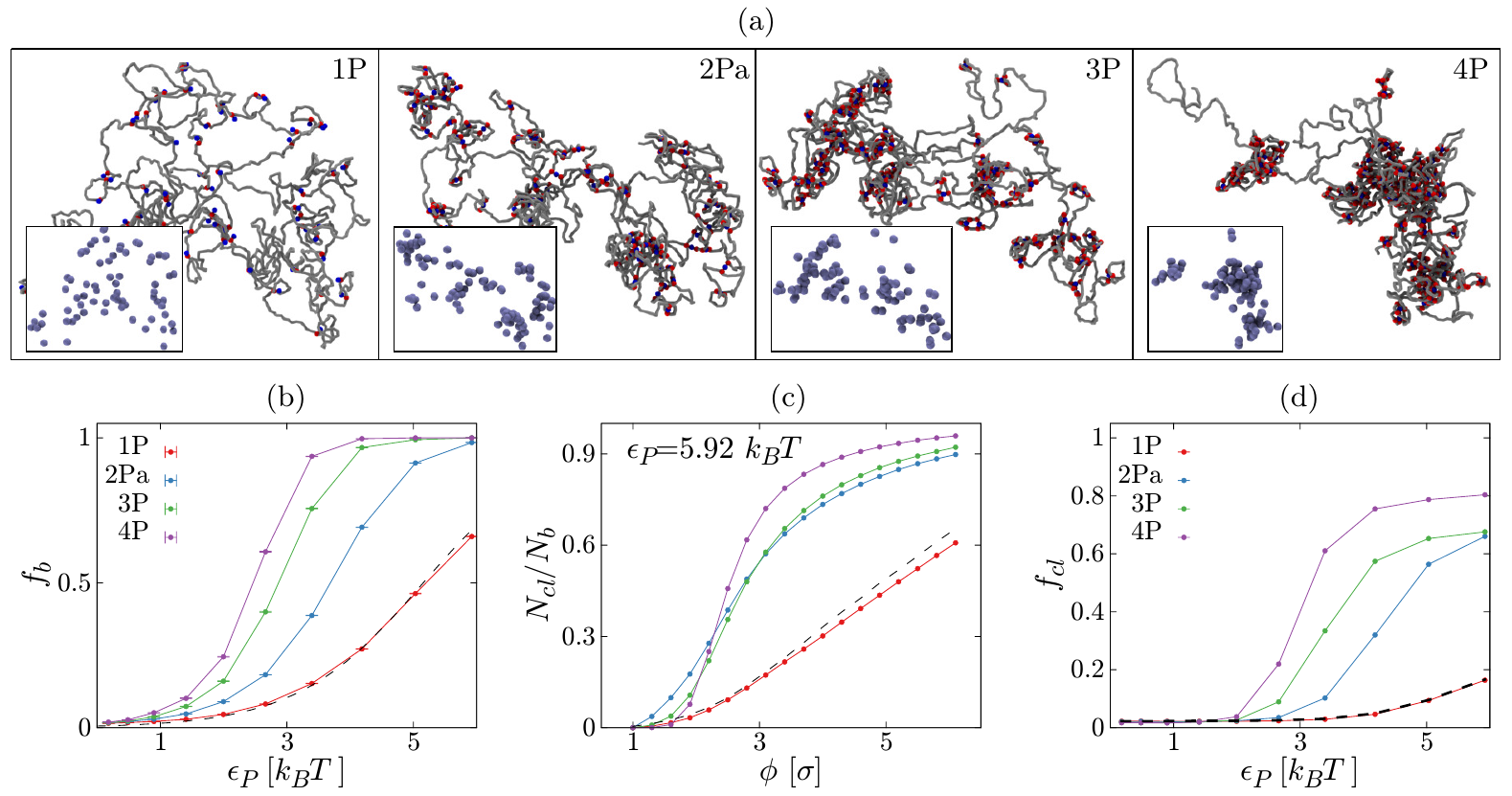}
\caption{The BIA for model proteins with different valence. Results from simulations of $N=100$ proteins interacting with an $L=2000$ bead polymer were performed in a square box of size $l_x=90~\sigma$. 
(a) Snapshots of typical configurations with different valence proteins for $\epsilon_P=5.92~k_BT$. For clarity, only the proteins bound to the polymer are shown (a protein is defined as bound if the separation between the protein central bead and a polymer bead is less than $1.7~\sigma$). Insets show the same configuration, but to give a visual impression of the degree clustering only the central beads of each protein are shown (with a larger diameter of 3.5~$\sigma$).
(b) The mean fraction of proteins bound to the polymer is plotted as a function of interaction energy. Points show data from simulations with the indicated protein model, and are an average over the last $10^5\tau$ of 5 independent simulation runs, with connecting lines as a guide for the eye. The dashed line shows a fit to the 1P data of the function given in the text (fit constant $A=173.5$). 
(c) Ratio of the number of proteins in clusters and the number of proteins which are bound to the polymer as a function of the cluster threshold $\phi$. The dashed line shows the level of clustering expected simply due to proteins binding randomly (non-cooperatively) to a polymer coil (see text and \anappendix{C}).
(d) Fraction of proteins which belong to clusters as a function of interaction energy (a protein belongs to a cluster if the centre of its central bead is within $\phi=3.5~\sigma$ of that of another protein). The dashed line again shows the expected clustering simply due to random binding (see text and \anappendix{C}).
\label{fig:BIAhigher}}
\end{figure*}

\subsection*{The BIA for patchy proteins with different valence}

In this section we present simulations of patchy proteins with different valence, and study the formation of clusters due to the BIA. We consider a system with $L=2000$ polymer beads and $N=100$ proteins of one type in a cube of size $l_x=90~\sigma$. The number of proteins is small compared to the number of polymer beads (protein binding sites) in order that discrete clusters will be detectable if they arise (compare the case below where the ratio $N/L$ is larger and the polymer is saturated by proteins). Some typical snapshots are shown in Fig.~\ref{fig:BIAhigher}(a). 

We first consider the fraction of proteins $f_b$ which are bound to the polymer on average at any one time [Fig.~\ref{fig:BIAhigher}(b)]. For the 1P model we can approximate the behaviour using a simple kinetic binding model, i.e., where unbinding occurs at some rate $k_{\rm off}\sim e^{-\epsilon_P/k_BT}$, and is independent of the number of proteins already bound. This leads to a steady state
\begin{equation}\label{eq:sig}
f_b = 1/(1+Ae^{-\epsilon_P/k_BT}),
\end{equation}
where $A$ is a constant (see \anappendix{B}). A fit is shown as a black dashed line in Fig.~\ref{fig:BIAhigher}(b).
For the other models, we find that at low energy the number of proteins bound grows with the protein valence. Since each patch can interact with the polymer separately, the total binding energy per protein is the sum of these contributions---thus we expect the mean residence time $k_{\rm off}^{-1}$ to increase with the number of patches. These curves do not fit well to a function of the form given in Eq.~(\ref{eq:sig}), but instead $f_b$ increases more steeply, suggesting cooperative binding (due to the BIA) which increases with valence.

To quantify clustering, we define a protein as belonging to a cluster if its centre bead is less than some threshold distance $\phi$ away from that of another protein (i.e., the smallest clusters contain two proteins). Rather than choosing an arbitrary value for $\phi$, we instead examined the number of proteins found to be in clusters, $N_{cl}$ as a function of $\phi$ for a high value of $\epsilon_P$. Since the different valence proteins bind the polymer to a different extent, in Fig.~\ref{fig:BIAhigher}(c) we plot $N_{cl}$ scaled by the mean number of bound proteins $N_{b}$. The multivalent proteins (models 2Pa, 3P, and 4P) show an initially steep increase with $\phi$, but this begins to plateau, suggesting clustering, as expected due to the BIA. The 1P proteins show a strikingly different behaviour: after an initial slow increase in $N_{cl}$ with $\epsilon_P$ the dependence becomes linear. We expect that this clustering is simply due to proteins binding randomly along the polymer coil---this can be confirmed using polymer configurations from a simulation without proteins, and choosing $M$ polymer beads at random to attach proteins to, before performing the same clustering analysis (dashed black line in Fig.~\ref{fig:BIAhigher}(c); $M$ is the number of proteins bound in the 1P simulation). See \anappendix{C} for further details. Based on Fig.~\ref{fig:BIAhigher}(c) we suggest that $\phi=3.5~\sigma$ is a reasonable threshold to determine clustering. Figure~\ref{fig:BIAhigher}(d) then shows the fraction of proteins involved in clusters $f_{cl}$ as a function of $\epsilon_P$ for each protein model. Consistent with the above discussion, the 1P proteins show a weak increase with $\epsilon_P$, while the other protein models show a steep increase and plateau. The curves become steeper (with clustering observed at lower energy) as the number of patches increases, indicating that the BIA effect becomes stronger with increasing valence. 

\begin{figure}[t!]
\centering
\includegraphics[width=0.47\textwidth]{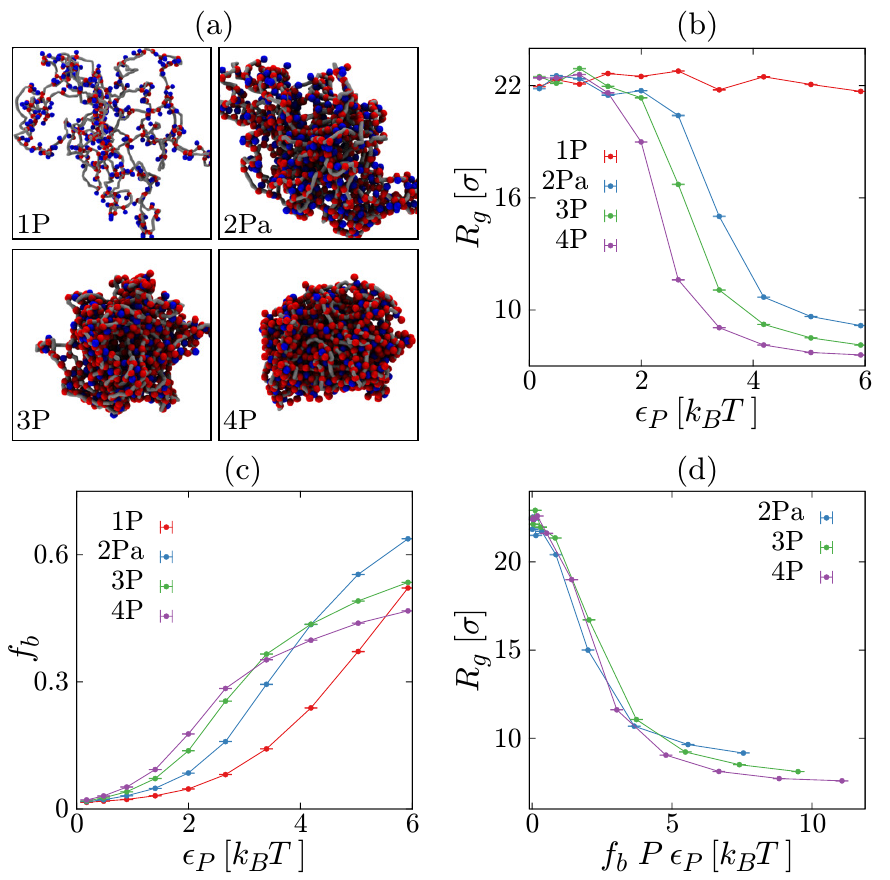}
\caption{Polymer compaction by patchy proteins with different valence. Results from simulations of $N=500$ proteins interacting with an $L=1000$ bead polymer in a square box of size $l_x=70~\sigma$.
(a) Snapshots of typical configurations with different valence proteins for $\epsilon_P=5.92~k_BT$. For clarity, only the proteins bound to the polymer are shown.
(b) The mean polymer radius of gyration is plotted as a function of interaction energy $\epsilon_P$. Points show an average over the last $10^5\tau$ of 5 independent simulations, with connecting lines as a guide to the eye.
(c) Similar plot showing the mean fraction of proteins bound to the polymer.
(d) The radius of gyration is plotted as a function of the mean total protein-polymer interaction energy, given by $f_bP\epsilon_P$, where $P$ is the number of patches (valence). The curves roughly collapse on top of each other.
\label{fig:compacthigher}}
\end{figure}

\subsection*{Polymer compaction by patchy proteins with different valence}

We next characterise the ability of the different protein models to compact a polymer. We perform simulations with a smaller $L=1000$ bead polymer and $N=500$ proteins in a simulation box of side $l_x=70~\sigma$; in this case there are sufficient proteins to compact the whole polymer. Typical snapshots are shown in Fig.~\ref{fig:compacthigher}(a). In Fig.~\ref{fig:compacthigher}(b) we plot the radius of gyration of the polymer, defined as
\[
R_g^2 = \frac{1}{L} \sum_{i=1}^{L} ( \mathbf{r}_i - \mathbf{r}_{c} )^2
\]
in order to asses polymer compaction. Here $\mathbf{r}_i$ is the position of polymer bead $i$, and $\mathbf{r}_c = (1/L) \sum_i \mathbf{r}_i$. We find that, as expected, no compaction is observed for the 1P model which cannot form bridges. The other models all show compaction when the protein-polymer interaction strength $\epsilon_P$ is large. For the 4P case a steep decrease in $R_g$ is observed at around $\epsilon_P\approx2~k_BT$; this behaviour is similar to the case of isotropic spheres~\cite{Nicodemi2009}, and reminiscent of the well studied polymer coil-globule transition for a poor solvent~\cite{deGennes1979,Bhattacharjee2013}. As the number of patches decreases, the decrease in $R_g$ shifts to larger $\epsilon_P$, indicating that proteins with fewer patches are less able to compact the polymer. 

Figure~\ref{fig:compacthigher}(c) shows the fraction of proteins which are bound to the polymer on average at any one time, $f_b$. This looks similar to the case with fewer proteins as shown in Fig.~\ref{fig:BIAhigher}(b). Again, at small $\epsilon_P$, $f_b$ increases with protein valence, and (except for the 1P case) there is a plateau at large $\epsilon_P$. This time, however, at large $\epsilon_P$ there is a clear trend for the higher valence proteins to show lower $f_b$ values---this is likely because the ``binding sites'' on the polymer are becoming saturated. 

Since for the different models, different numbers of proteins are bound for a given value of $\epsilon_P$, it is useful to consider the level of polymer compaction as a function of the total protein-polymer interaction energy, $f_bP\epsilon_P$, where $P$ is the number of patches per protein (valence). Scaling the binding energy in this way collapses the $R_g$ curves on top of each other for the 2P-4P models [Fig.~\ref{fig:compacthigher}(d)]. One can rationalise this by considering that to compact the polymer to the same extent one would need twice as many 2Pa proteins as 4P proteins for the same $\epsilon_P$ (a cluster of two 2Pa proteins looks much like a single 4P protein, etc.).

\begin{figure*}[ht!]
\centering
\includegraphics{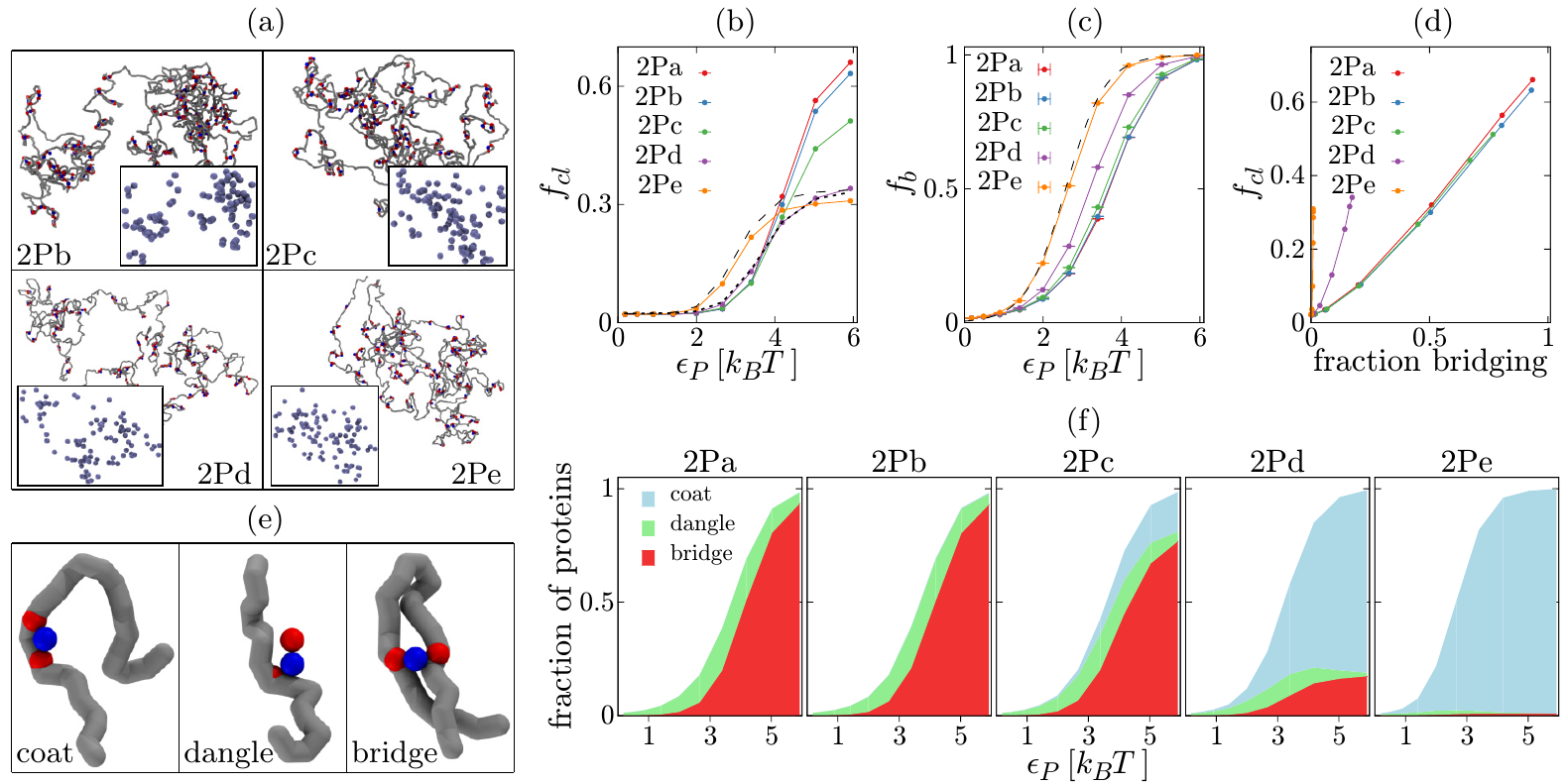}
\caption{\label{fig:2patchClust} 
Bridging-induced clustering of valence-2 model proteins with different shape ($N=100$ proteins interacting with an $L=5000$ bead polymer in a cube of size $l_x=90~\sigma$). (a) Snapshots of typical configurations for the 2Pb-e models from simulations where $\epsilon_P=5.92~k_BT$. As before, only proteins which are bound to the polymer are shown. Insets show the same configuration, but only the central beads of each protein are shown (with a larger diameter of 3.5~$\sigma$). (b) The fraction of proteins which are found in clusters on average is plotted as a function of $\epsilon_P$ for the different models. Points show an average over the last $10^5\tau$ of 5 independent simulations, with connecting lines as a guide to the eye. The dashed line shows the level of clustering expected simply due to random binding, for the level of binding observed in the 2Pe case. The dotted line shows similar, but for the level of binding observed in the 2Pd case (see \anappendix{C}).
(c) Similar plot showing the fraction of proteins which are which are bound on average. The dashed line shows the function given in Eq.~(\ref{eq:sig}), but with the replacement $\epsilon_P\rightarrow2\epsilon_P$ (and $A=173.5$ as before, see text).
(d) The fraction of proteins found in clusters is plotted, this time as a function of the number of proteins which are bound in the ``bridging'' mode. For the 2Pc-e models the curves collapse on top of each other.  
(e) Snapshots showing the three different binding modes, ``coating'', ``dangling'', and ``bridging'', as detailed in the text.
(f) A plot for each protein model shows the proportion of the $N=100$ proteins binding in each mode as a function of $\epsilon_P$. Curves are stacked on top of each other, so e.g., the height of the green region gives the proportion of proteins which bind in the dangling mode. }
\end{figure*} 

\subsection*{Effect of shape on the BIA for valence-2 patchy bridges}

\vspace{-2mm}

Above we considered model proteins where the patches were placed equidistantly around the central bead. Of course, in real proteins such an arrangement of DNA binding domains may  not the be case. To systematically study the effect of different protein shapes, in this section we consider valence 2 proteins and vary the angle between the patches. Specifically, we consider the five model proteins as shown in the bottom row of Fig.~1(b). Real examples of proteins which might have a similar arrangement of binding domains include the eukaryotic protein HP1 (which is thought to form dimers either in an elongated or a folded state, depending on post-translational biochemical modifications~\cite{Larson2017}), and the bacterial protein H-NS (which is thought to adopt elongated or folded shapes depending on salt concentration or the presence of other ligands). Both of these examples could therefore, under different conditions, be similar to either the 2Pa or 2Pe models. 

To study the formation of clusters via the BIA, we again consider a system with an $L=2000$ bead polymer and $N=100$ proteins with $l_x=90~\sigma$; simulation snapshots are shown in Fig.~\ref{fig:2patchClust}(a). In Fig.~\ref{fig:2patchClust}(b) we plot the fraction of proteins found in clusters for each of the different models (using a threshold separation $\phi=3.5~\sigma$, as before). The 2Pb protein model behaves very similarly to 2Pa---as the binding energy increases, there is a sharp increase in clustering at around $\epsilon_P=4~k_BT$. For the 2Pe model, where the DNA binding patches are both on the same side of the molecule, we see quite different behaviour---the proteins start to cluster at smaller $\epsilon_P$, and by $\epsilon_P\sim4~k_BT$ the level of clustering starts to plateau (at a much lower level than in reached, e.g., by the 2Pa proteins). By measuring the number of proteins bound, we can again estimate the level of clustering which would be observed just due to random binding to a polymer coil in the absence of any cooperative effects [black dashed line in Fig.~\ref{fig:2patchClust}(b)]. We find that the clustering observed in the 2Pe case is approximately at this level, indicating that the BIA is not in effect (at high energy for 2Pe, $f_{cl}$ is actually lower than expected due to random binding, suggesting that these model proteins have a larger excluded volume when bound than taken into account by our random binding estimate -- see \anappendix{C}). The 2Pd model also only shows the level of clustering expected from random binding to the polymer [black dotted line in Fig.~\ref{fig:2patchClust}(b)]. The 2Pc curve shows a behaviour somewhere between the two.

Figure~\ref{fig:2patchClust}(c) shows how the level of protein binding $f_b$ depends on $\epsilon_P$. For all of the models we observe an ``S-shaped'' curve. One might naively expect that, since we do not observe clustering for the 2Pe proteins, there would be no cooperative effects, and so, e.g., $f_b$ would pass through 0.5 at larger $\epsilon_P$. In fact, the opposite is the case. The 2Pe curve crosses $f_b=0.5$ at \textit{lower} $\epsilon_P$, while the 2Pa and 2Pb curves virtually overlap, crossing $f_b=0.5$ at larger $\epsilon_P$. The curves for the other models sit somewhere in between. These observations can be rationalised by considering that the valence-2 proteins can bind the polymer in several different `modes' [Fig.~\ref{fig:2patchClust}(e)]: either just one of the patches binds the polymer, or they both do. In the latter case, patches can bind two beads which are nearby along the polymer contour, or two beads which are far apart; we describe these as ``coating'' and ``bridging'' bonds respectively. Specifically, if a protein binds polymer beads $i$ and $j$, we say it is coating if $|i-j|<3$, otherwise it is bridging. When only one patch binds, we call this ``dangling''. Figure~\ref{fig:2patchClust}(f) shows, for each protein model, the fraction of proteins which bind in each mode as a function of $\epsilon_P$. For the 2Pa and 2Pb models we find that most proteins bind in the bridging mode, with a small fraction ``dangling''. As the angle between the patches decreases (progressively through the 2Pc, 2Pd and 2Pe models), a larger proportion of proteins bind in the coating mode---in the 2Pe case nearly all proteins bind in this way. This explains the observed trends for clustering: the BIA is only in effect when proteins can form bridges. In fact, if we plot $f_{cl}$ as a function of the fraction of proteins binding in the bridging mode [Fig.~\ref{fig:2patchClust}(d)], the curves for the 2Pa, 2Pb and 2Pc curves collapse on top of each other. The 2Pe proteins actually behave like valence-1 proteins, but since they have two binding patches, their polymer interaction effectively has double the energy (the black dashed line in  Fig.~\ref{fig:2patchClust}(c) shows the curve given by Eq.~(\ref{eq:sig}) with the replacement $\epsilon_P\rightarrow2\epsilon_P$ using the constant $A$ found from the fit to the 1P proteins).

We also examined how protein binding affected bending of the polymer by measuring the mean angle $\langle\theta_i\rangle$ between consecutive polymer segments (defined as in Eq.~(\ref{eq:Vbend})). This revealed that the 2Pe proteins tend to slightly stiffen the polymer (reducing $\langle\theta_i\rangle$), whereas the 2Pd proteins tend to induce bends (increasing $\langle\theta_i\rangle$). As expected, binding of proteins which readily form bridges (2Pa-c), and therefore create loops, also leads to an increased bending. These effects are all very small with $\langle\theta_i\rangle$ changing by less than 0.5\% with respect to a polymer coil in the absence of proteins; the effect on the persistence length as measured from the decay in bond-bond correlations ($\langle \mathbf{\hat{t}}_i\cdot\mathbf{\hat{t}}_{i+s} \rangle_i = e^{-s/L_p}$) is smaller than the error in the measurement. Previous work~\cite{Rappaport2008,Rappaport2009,Zhang2010} has shown that proteins which induce bending or a stiffening of a polymer can experience mechanically induced cooperative binding, but these effects appear to be too weak to lead to clustering in the case of the model proteins studied here.

\begin{figure}[ht!]
\centering
\includegraphics[width=0.48\textwidth]{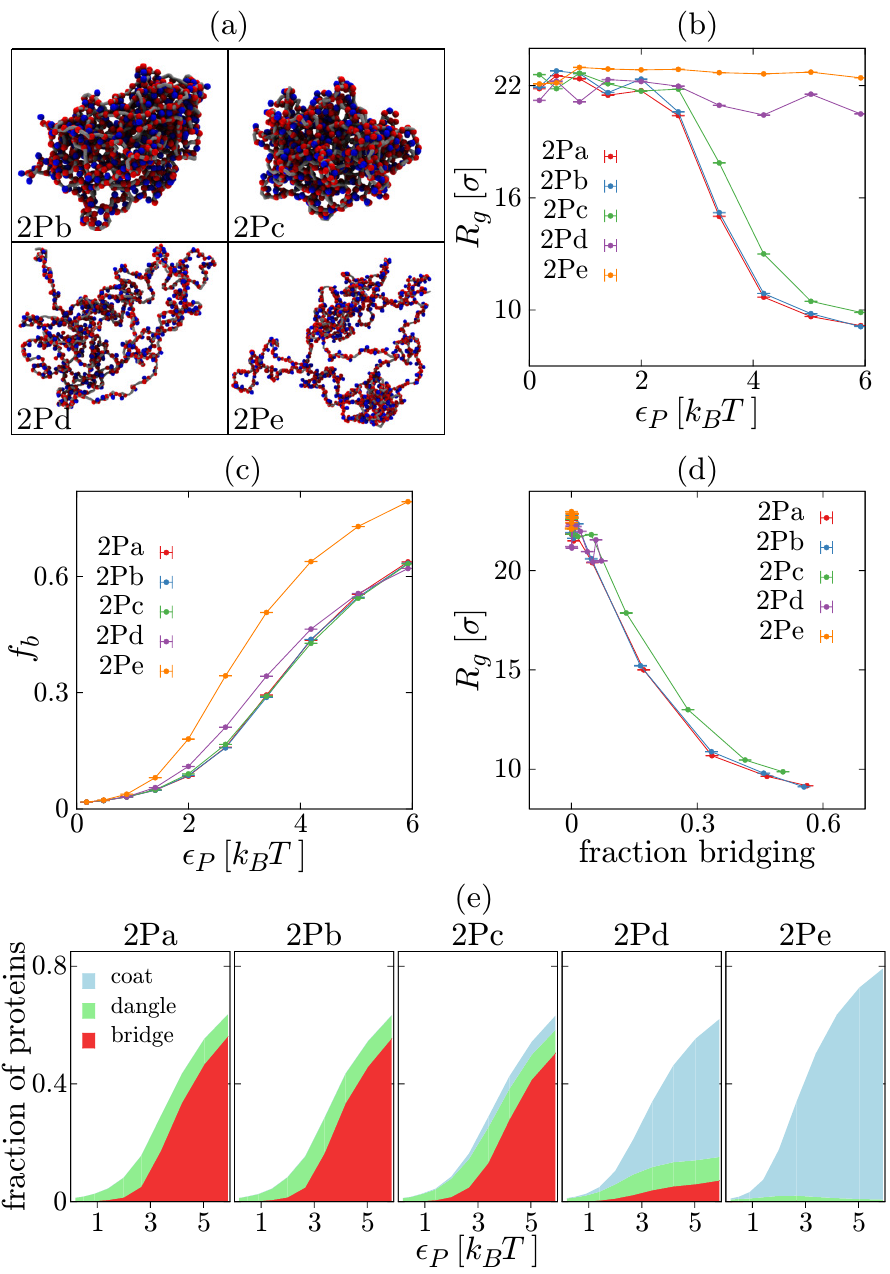}
\caption{\label{fig:2patchCompact}
Polymer compaction by valence-2 model proteins ($N=500$ proteins interacting with an $L=1000$ bead polymer in a cube of size $l_x=70~\sigma$). (a) Snapshots showing typical equilibrium configurations in simulations where $\epsilon_P=5.92~k_BT$. As before, only proteins which are bound to the polymer are shown.
(b) The radius of gyration is plotted as a function of $\epsilon_P$ for each model protein. Points show an average over the last $10^5\tau$ of 5 independent simulations, with connecting lines as a guide to the eye.
(c) Similar plot showing the fraction of proteins bound on average. 
(d) The radius of gyration is plotted as a function of the mean number of proteins bound in the ``bridging'' mode.
(e) A plot for each protein model shows the fraction of the $N=500$ proteins which bind in each of the bridging, dangling or coating modes [see Fig.~\ref{fig:2patchClust}(e)] as a function of $\epsilon_P$. }
\vspace{-1cm}
\end{figure}

\subsection*{Effect of shape on polymer compaction for valence-2 patchy bridges}

\vspace{-2mm}

Finally, in Fig.~\ref{fig:2patchCompact} we present results from simulations with $N=500$ valence-2 proteins interacting with an $L=1000$ bead polymer ($l_x=70~\sigma$), and again examine the ability of the proteins to compact the polymer. The radius of gyration decreases with $\epsilon_P$ only for the 2Pa-c protein models. The 2Pa and 2Pb curves overlap, indicating that these proteins can compact the polymer to a similar extent ($R_g$ tends to be slightly higher for 2Pc).
The fraction of proteins bound as a function of $\epsilon_P$ [Fig.~\ref{fig:2patchCompact}(c)] shows similar trends as the case with fewer proteins---the 2Pe proteins bind more readily. Fig.~\ref{fig:2patchCompact}(e) shows the proportion of proteins which are binding in each mode as a function of $\epsilon_P$; again the behaviour is very similar to that in the previous section, with the proteins more likely to coat rather than bridge as the angle between patches decreases. The proportion which bind in the ``dangling'' mode is slightly larger at higher $\epsilon_P$ values than in the case with fewer proteins (this is likely due to ``binding sites'' on the polymer becoming saturated). Figure~\ref{fig:2patchCompact}(d) again shows $R_g$, but this time as a function of the fraction of proteins bound in the bridging mode: there is a rough collapse of the curves for the different proteins. Together these results suggest that again, the effect of protein shape mainly comes down to the ability of each model to form bridges.

\section*{Discussion and Conclusions}

\vspace{-3mm}

In this work we have further investigated the bridging-induced attraction (BIA), a mechanism which was uncovered in simulations of sphere-polymer mixtures~\cite{Brackley2013}. Previously, we showed that spheres with an isotropic attractive interaction with a polymer tend to spontaneously form clusters, even in the absence of attractive sphere-sphere interactions. These clusters grow and coarsen until there is a single large cluster~\cite{BrackleyNAR}. In this way the BIA leads to a bridging-induced phase separation. Spheres are the simplest model for DNA or chromatin binding proteins (or protein complexes) which can bind in multiple places simultaneously in order to form molecular bridges. In the present work we study in more detail the effect of protein valence on the BIA by using simple ``patchy particles'' to model multivalent proteins.

We found that all model proteins with valence $\geq2$ form clusters, but that the effect is stronger for higher valence. The fraction of proteins involved in ($\geq 2$ protein) clusters, $f_{cl}$ shows an ``S-shaped'' dependence on the protein-polymer interaction energy. The fraction of proteins bound on average $f_b$ also shows an S-shaped curve, as expected. In comparison to valence-1 proteins (or a simple kinetic binding model), the $f_b$ curves for multivalent proteins are steeper and, for example, cross $f_b=0.5$ at lower energy as the valence increases---this indicates cooperative binding. This is consistent with the action of the BIA, where after an initial bridge forms, subsequent bridges forming in the same place act to stabilise the first.

In our previous work on isotropic spheres~\cite{BrackleyEphemeral} we found that bridging-induced clusters coarsen, growing in size until only one large cluster remains at equilibrium (perhaps with smaller clusters which form transiently). These dynamics can be very slow compared to the duration of a typical simulation; two possible coarsening mechanisms are cluster coalescence (which requires large, slowly diffusing clusters to come into contact, and can be hindered by intervening unbound polymer), or shrinking of small clusters at the expense of large ones (where sufficient time must be allowed for proteins to unbind and rebind the polymer). It would be interesting to understand whether cluster coarsening also occurs with the lower valence protein models. Particularly, is the equilibrium state fully phase separated, or a microphase separation? In practice it was not possible to detect cluster coarsening in the relatively small and short simulations presented here. While it is beyond the scope of the present work, we propose that in the future, longer simulations (perhaps employing enhanced sampling techniques) of larger systems in which many small clusters initially form could be used to study cluster coarsening. In other systems (e.g. a phase separating binary fluid mixture) coarsening is associated with interfacial tension; in the present system, cluster surface tension is provided by ``unbound'' patches and polymer beads at the edges of clusters~\cite{Broedersz2014}. One might speculate that with the lowest valence proteins, it will be easier to satisfy all protein bounds and so the surface tension will be lower, which could lead to an arrest of coarsening (though this picture clearly neglects the entropic contribution of the polymer configuration).

For the case of valence-2 proteins we found that the BIA also depends on the protein shape. Specifically, for our simple patchy proteins we varied the position of the two patches; starting from a model where they were on opposite sides of the protein ``central bead'' (model 2Pa) we progressively reduced the angle between the patches across five models, with the final model having two adjacent patches on one side of the central bead (model 2Pe). The level of clustering was higher for a larger angle between the patches. This effect can be traced back to the ability of the proteins to form bridges. For example, the 2Pe model proteins much more readily bind the polymer in a ``coating'' arrangement---if no bridge is formed, the BIA will not be in effect. This is confirmed by a plot of the fraction in clusters $f_{cl}$ as a function of the fraction of proteins forming bridges, where (aside from clustering which occurs simply due to localisation of proteins to the polymer, which dominates for the 2Pd-e models) the curves collapse on top of each other [Fig.~\ref{fig:2patchClust}(d)].

As noted above, there are several examples of proteins which are thought to have structures similar to our valence-2 model proteins in terms of the layout of their DNA binding domains. This includes the histone-like nucleoid structuring protein (H-NS), which has also previously been studied using simple coarse-grained simulations~\cite{Joyeux2013,Joyeux2018}. These proteins readily form dimers (though larger oligomers are also observed) which adopt either an elongated (2Pa-like) or folded (2Pe-like) shape depending on salt concentration or the presence of other factors. An interesting system which could be studied further in the future is a mixture of different model proteins, or proteins which can switch between two states/shapes~\cite{Joyeux2018}. One could consider proteins which transition thermodynamically between two states, i.e., there is an energy barrier between the states, which can be surmounted by thermal fluctuations (and polymer binding might act to stabilising the protein against switching state). Or, one could consider active (non-equilibrium) switching, as in Ref.~\cite{BrackleyEphemeral}, where the proteins switch between states independently of thermal fluctuations in a way that breaks detailed balance. The latter might represent a protein which undergoes refolding after, e.g., post-translational chemical modification or interaction with ATP.

When there is an excess of proteins (compared to the number of polymer beads, or ``binding sites''), rather than clusters we observe a polymer collapse. This is reminiscent of the coil-globule transition which is observed for a polymer in a poor solvent~\cite{deGennes1979,Bhattacharjee2013}, and has previously be studied for the case of isotropic spheres~\cite{Nicodemi2009,Barbieri16173}. Here we studied the ability of the different model proteins to compact a polymer by measuring the radius of gyration; we find that only multivalent proteins can collapse/compact the polymer, and as valence increases the ``compacting ability'' of the proteins also increases [the decrease in $R_g$ occurs at lower values of $\epsilon_P$, Fig.~\ref{fig:compacthigher}(b)]. The $R_g$ curves for the different models collapse on top of each other if we plot as a function of $f_bP\epsilon_P$ [Fig.~\ref{fig:compacthigher}(d)], indicating that in this regime the compaction depends on the total protein-polymer attraction energy. 
In regards to the polymer compacting ability of the valence-2 proteins, we found that this again depended on the propensity of the proteins to form bridges. Protein models which tend to bind in a coating mode (models 2Pd and 2Pe) did not induce polymer collapse. Consistent with this, plotting the radius of gyration as a function of the number of proteins binding in the bridging mode gave curves for the different models which roughly overlap.

These results are highly relevant for understanding the physics of the formation of genome-associated protein foci in cells. While many proteins only have a single DNA binding domain, these often function within large complexes. For example, transcription typically involves RNA polymerase, transcription factors, and co-activating factors such as the mediator complex. In recent years, an idea which has gained much traction in the molecular biology literature is that protein foci form as micro-phase separated liquid-like droplets mediated by protein-protein interactions~\cite{Hyman2014,Mir2019,Strom2019}. That view prompts a number of questions, including, ``\textit{What prevents such droplets from coarsening?}'', ``\textit{How are droplets `pinned' at the relevant genomic location?}'', and ``\textit{Do active biochemical processes play a role?}. The BIA offers an alternative (or, more likely, complementary) mechanism for cluster formation. It will be interesting in the future to study the interplay between protein-chromatin and protein-protein interactions, as well as the effect of non-equilibrium processes. The physics of such systems likely plays an important role in the formation of many different protein structures within the nucleus, impacting on genome regulation in both healthy and disease states.

\vspace{-3mm}

\subsection*{Appendix A: Details of patchy interactions}

\vspace{-2mm}

In this appendix we further detail two aspects of the patch-polymer interactions. 

First, it is intended that for all of the patchy protein models, each patch can only make one protein-polymer bond at a time---i.e., that a patch itself cannot form a bridge. To ensure this, in all simulations we counted the number of polymer beads that each patch interacts with (an interaction is defined as the patch-polymer centre-to-centre separation is less than the $V_{\rm PATCH}(r)$ cut off $r_c=0.9~\sigma$). We found that it was possible for a patch to sit between two consecutive polymer beads ($i$ and $i+1$), and we did not count this as bridging. The relative occurrence of patches interacting with a single polymer bead or with two consecutive beads depends on the interaction energy, and was observed in over 97\% of patch-polymer interaction events. Occasionally, a patch was found to be interacting with 3 polymer beads, but this was usually ($\sim2.5\%$ of all interaction events) with three consecutive beads ($i$,$i+1$,$i+2$); this type of interaction induces a bend in the polymer. Instances of a patch binding more than one non-consecutive polymer bead (true bridging by a patch) accounted for less than 0.5\% of patch-polymer interaction events across all simulations.

\begin{figure}
\centering
\includegraphics[width=0.48\textwidth]{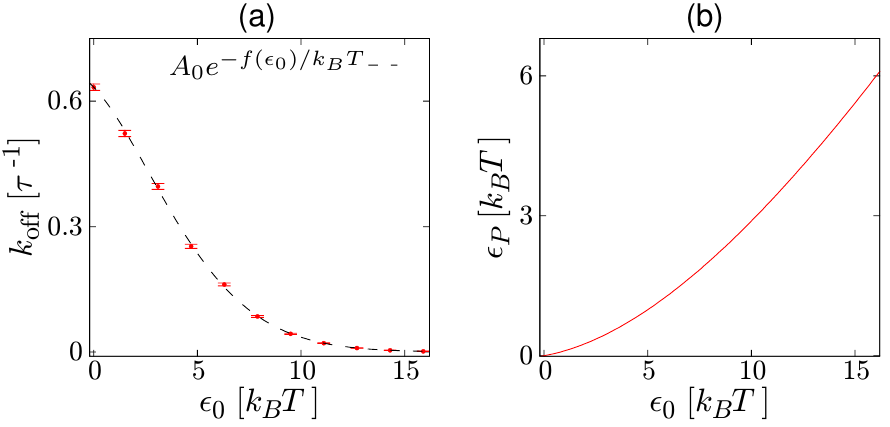}
\caption{\label{fig:energymapping}
Plots showing the energy mapping function $\epsilon_P=f(\epsilon_0)$ obtained from fitting. (a) Points show the measured unbinding rate $k_{\rm off}$ from simulations with a single model 1P protein and an $L=500$ bead polymer, for different values of the input energy $\epsilon_0$. The functional dependence on the actual patch-polymer interaction energy $\epsilon_P$ given in the text is assumed, and the function $f(\epsilon_0)$ is obtained from a fit (dashed line). (b) The function $f(\epsilon_0)$ is plotted directly.
}
\end{figure}

Second, we measured from the simulations the actual patch-polymer interaction energy $\epsilon_P$ for a given interaction strength setting $\epsilon_0$. These two energies are not equal for several reasons: (i) due to the functional form of $V_{\rm PATCH}(r_{ij})$ as given in Eq.~(\ref{eq:Vpatch}), the energy minimum is shifted; (ii) though the minimum of this function is at $V_{\rm PATCH}(r_{ij}=0)$, due to, e.g., steric interactions between the protein ``central bead'' and the polymer, it may not be possible for the patch-polymer bead separation to actually reach zero; (iii) as noted above, a patch can sit between two consecutive polymer beads, so the total interaction energy will be the sum of two contributions.
To accurately get a mapping function $\epsilon_P=f(\epsilon_0)$, we performed simulations with different $\epsilon_0$ for a single model 1P protein and an $L=500$ bead polymer, and measured the mean residence time for the protein on the polymer $\tau_{\rm res}=k_{\rm off}^{-1}$ as a function of $\epsilon_0$. Assuming that the unbinding rate is given by
\[
k_{\rm off} = A_0 e^{-\epsilon_P / k_BT },
\]
and obtaining $A_0$ from a simulation with $\epsilon_0$=0, we can obtain $\epsilon_P$ from the measured $k_{\rm off}$. We obtain the mapping function by fitting (Fig.~\ref{fig:energymapping}); the simplest function which gives a good fit to the data is $\epsilon_P=f(\epsilon_0)=C_1\epsilon_0 -C_2(1-e^{-\epsilon_0 /C_3})$ with $C_1=0.7915$, $C_2=9.675~k_BT$ and $C_3=13.627~k_BT$. Reassuringly the function becomes linear at large $\epsilon_0$.

\subsection*{Appendix B: Predicted binding fraction for valence-1 proteins}

\vspace{-2mm}

One can predict the fraction of valence-1 proteins which bind in steady state from a simple kinetic binding model. The rate of change of the bound fraction $f_b$ is
\begin{equation}\label{eq:simplekinetic}
\frac{df_b}{dt} = -k_{\rm off} f_b + k_{\rm on} (1-f_b),
\end{equation}
where $k_{\rm on}$ is the rate which proteins bind, and the unbinding rate $k_{\rm off}$ has the usual exponential dependence on the energy $k_{\rm off}\sim e^{-\epsilon_P/k_BT}$. Setting the left hand side to zero to obtain the steady state gives
\[
f_b = \frac{1}{1+Ae^{-\epsilon_P/k_BT}},
\]
with $A$ constant. Equation~(\ref{eq:simplekinetic}) assumes that there are unlimited binding sites on the polymer; this assumption could be relaxed by, e.g, including a factor $(L-f_b)$ in the $k_{\rm on}$ term. Equation~(\ref{eq:simplekinetic}) can also approximate the 2Pe protein case by assuming that all proteins bind in the ``coating'' mode, and adjusting the energy to account for the interaction of both patches (replacing $\epsilon_P\rightarrow2\epsilon_P$). The cooperative binding of the other protein types would effectively lead to non-linearities in $f_b$ in the $k_{\rm off}$ term.

\vspace{-4mm}

\subsection*{Appendix C: Protein clustering due to random binding on a polymer coil}

\vspace{-2mm}

If proteins bind at random positions on a polymer coil, then even in the absence of any cooperative binding effects (e.g., the BIA) a low level of clustering will still be observed. That is to say, when, e.g., $M$ out of $N$ proteins become localised to a polymer, this is enough to lead to clusters depending on the choice of clustering threshold $\phi$ and the size of the polymer. To quantify the level of clustering which would be observed due to this effect, we ran a set of simulations of a polymer ($L=2000$ confined in a $l_x=90~\sigma$ cubic box) in the absence of proteins. Using polymer coil configurations from this simulation, we can attach $M$ ``virtual proteins'' to random positions on the coil (assuming that binding a protein at bead $i$ prevents further binding at beads $i-1$ and $i+1$). We also position the remaining $N-M$ ``virtual proteins'' randomly within the simulation box. We then measure the level of clustering of these for a given value of $\phi$. 

The black dashed line in Fig.~\ref{fig:BIAhigher}(c) was generated by taking the number of bound proteins from the 1P simulation with $\epsilon_P=5.92~k_BT$ as the value of $M$, and performing the above procedure for different values of $\phi$. The black dashed line in Fig.~\ref{fig:BIAhigher}(d) was similarly generated, but this time for fixed $\phi=3.5~\sigma$ and values of $M$ taken as the number of proteins bound in the 1P simulations for each different $\epsilon_P$ value.

The black dashed line in Fig.~\ref{fig:2patchClust}(b) is obtained by performing the above procedure with $\phi=3.5~\sigma$ using values of $M$ from the number of proteins bound in the 2Pe simulations for each different $\epsilon_P$ value. At large $\epsilon_P$, this prediction of clustering due to random non-cooperative binding is slightly higher than observed in the 2Pe simulations---we expect this is because 2Pe binding at polymer bead $i$ sterically excludes binding at more surrounding beads than accounted for in the ``virtual protein'' procedure detailed above. 

The black \textit{dotted} line in Fig.~\ref{fig:2patchClust}(b) is obtained using values of $M$ from the number of proteins bound in the 2Pd simulations.

\vspace{-5mm}

\subsection*{Acknowledgements}

\vspace{-2mm}

We acknowledge the European Research Council for funding (Consolidator Grant THREEDCELLPHYSICS, Ref. 648050).

\vspace{-5mm}


\balance

\end{document}